# Oxygen Isotope Effect Resulting from Polaron-induced Superconductivity in Cuprates

S. Weyeneth

*Physik-Institut der Universität Zürich, Winterthuerstrasse 190, CH-8057 Zürich, Switzerland*

**Abstract**

The planar oxygen isotope effect coefficient measured as a function of hole doping in the Pr- and La-doped $YBa_2Cu_3O_7$ (YBCO) and the Ni-doped $La_{1.85}Sr_{0.15}CuO_4$ (LSCO) superconductors quantitatively and qualitatively follows the form originally proposed by Kresin and Wolf [1], which was derived for polarons perpendicular to the superconducting planes. Interestingly, the inverse oxygen isotope effect coefficient at the pseudogap temperature also follows the same formula. These findings allow the conclusion that the superconductivity in YBCO and LSCO results from polarons or rather bipolarons in the $CuO_2$ plane. The original formula, proposed for the perpendicular direction only, is obviously more generally valid and accounts for the superconductivity in the $CuO_2$ planes [2].

Copper oxides are the only compounds that show superconductivity at $T_c$ above the boiling point of nitrogen. When not doped, these materials are antiferromagnetic insulators due to the strong correlation splitting of the $d_{x^2-y^2}$-type band into an upper empty and a lower occupied Mott–Hubbard band. When the latter is doped and holes are present, very high $T_c$'s are found. A two-decade-long controversy resulted from the question as to whether the superconducting phenomenon still results from the electronic correlations present or whether lattice dynamics play a role. In favor of the latter, Kresin and Wolf wrote a colloquium [1] in which they review couplings from an extended lattice to phonons, plasmons, exitons and polarons, both theoretically and experimentally. For the case of polarons, i.e., a local vibronic lattice deformation, they quote a formula for the oxygen isotope coefficient $\alpha(n)$, originally derived by them [1, 3] in 1994 for lattice distortions along the $c$-direction:

$$\alpha(n, T_c) = \gamma(n) \frac{n}{T_c(n)} \frac{\partial T_c(n)}{\partial n}. \quad (1)$$

Here $n$ denotes the superconducting carrier density. The parameter $\gamma(n)$ entering Eq. (1) depends only weakly on $n$.

Equation (1) resembles the formula proposed earlier by Schneider and Keller, derived by analysing the universal proproties of cuprates [4, 5]. Already there, an important ingredient for mapping the experimentally obtained $\alpha(n)$ was the detailed $T_c(n)$ dependence. In the original report by Kresin and Wolf [1], the applicability of the model is discussed on the basis of the early data for $\alpha(n)$. The result of their fitting of Eq. (1) to the data yielded $\gamma = 0.13$. However, it is not really discussed how and in which manner the gradient of $T_c(n)$ is incorporated into Eq. (1) during fitting.

Alex Müller and me used to discuss occasionally about scientific articles being published shortly after the discovery of superconductivity in cuprates [6]. At some afternoon, he mentioned the above work of Kresin and Wolf [1, 3] and raised the idea of using Eq. (1) with more recent data in order to investigate weather also lattice distortions in the ab-plane follow the same trend. This idea was the starting point of the here revisited work which we published some years ago [2].

Experimentally, it was observed by Zech et al. [7] that site-selective oxygen substitution $^{16}O \rightarrow {}^{18}O$ in optimally doped $YBa_2Cu_3O_7$ (YBCO) results in more than 80% of the oxygen isotope effects being due to the planar oxygen. This is because the hole density present was shown to reside mainly in the planar $CuO_2$ [8]. Accordingly, the carrier density is mainly planar $n_{//}$. Note that for the perpendicular direction $n_\perp = 0$ and therefore also $\alpha_\perp = 0$.

Various groups [9–12] carefully studied $\alpha_{//}(n)$ for YBCO and doped La$_{1.85}$Sr$_{0.15}$CuO$_4$ (LSCO) as shown in Fig. 1, taken from Ref. [13]. In it, $\alpha_{//}(n)$ is compared with the results of a vibronic theory. The latter comprises two electronic bands present in the cuprates, a lower $t$–$J$-type $d$-band and a nearby, higher-lying $s$-band coupled by a linear vibronic term [8, 13, 14]. Within this model, the oxygen isotope effects on $T_c$ result from considering in-plane bands with first $t_1$, second $t_2$, third neighbor $t_3$, and interplanar hopping integrals $t_4$ [8, 13]. The polaronic renormalizations of the second-nearest-neighbor hopping integral $t_2$ and the interplanar hopping term $t_4$ yield the correct trend for $\alpha_{//}(n)$ close to optimal doping. From Fig. 1 it can be seen that the agreement is rather good at and below optimal doping, but clearly deviates for very low doping. Being based on a mean-field theory, this may be expected because at very low doping individual polarons exist. A result of that theory is that the local coupling mode is of Jahn–Teller-type $t_2/t_4$ motion. From this we may infer that for a polaronic scenario the formal $\alpha(n)$ [see Eq. (1)] may also be valid for the planar polarons. Thus we identify in Eq. (1) $n = n_{//}$ and $\alpha(n) = \alpha_{//}(n)$, as described in the next paragraph.

The dependence of $T_c(n)$ can be approximately parameterized according to early work [14-17] as

$$\frac{T_c(n)}{T_{c,\max}} = 1 - A\left(\frac{n}{n_{\max}} - 1\right)^2. \qquad (2)$$

$T_{c,\max}(n)$ denotes the maximum transition temperature of a given family, $n_{\max}$ the corresponding carrier density, and $A$ an empirical constant found to be $\approx$ 2–3 [14-17]. In the inset of Fig. 2 the dependence in Eq. (1) is presented. Combining Eqs. (1) and (2), we obtain for the generic formula of $\alpha_{//}(n)$ in terms of $z = n/n_{\max}$ and $t = T_c/T_{c,\max}$:

$$\alpha_\parallel(z) = 2\gamma(n)\frac{z(z-1)}{(z-1)^2 - A^{-1}} \qquad (3)$$

and within the underdoped region:

$$z = 1 - \sqrt{\frac{1-t}{A}}. \qquad (4)$$

This result can be compared with experimental data for $\alpha_{//}(n)$ from the literature. In Fig. 2, we depict data for $\alpha_{//}(n)$ as a function of $T_c/T_{c,\max}$ for Y$_{1-x}$Pr$_x$Ba$_2$Cu$_3$O$_{7-\delta}$ [9, 10], for YBa$_{2-x}$La$_x$Cu$_3$O$_{7-\delta}$ [11], and for La$_{1.85}$Sr$_{0.15}$Cu$_{1-x}$Ni$_x$O$_4$ [12]. Clearly, Eq. (3) describes the universal dependence of the measured $\alpha_{//}(n)$ rather well, assuming $\gamma_{//}(n)$ to be constant. By fitting Eqs. (3) and (4) to the data, we derive a value of $A$ = 2.5(3), in good agreement with literature values [14-17], and $\gamma$ = 0.146(8), slightly larger than the value of $\gamma$ = 0.13 reported in Ref. [1]

based on early results for $\alpha(n)$. Globally, this calculation is similar to those previously discussed in Refs. [1] and [4].

The agreement between the measured oxygen isotope effect $\alpha_{//}(n)$ as a function of doping $n$ and the curve calculated with Eqs. (3) and (4), but with a constant parameter $\gamma$, is really remarkable. From optimal doping $n_{max}$ down to near-vanishing superconductivity, where the measured $\alpha_{//}(n)$ exhibits a characteristic upturn, the data follow the curve obtained with Eq. (1) very well. This allows the conclusion that this simple expression is more generally valid than only for polarons axed along the crystallographic *c*-direction, for which it was originally obtained [1, 3]. It quantitatively yields the correct behavior *for polarons sited in the CuO$_2$ plane.* Now the vibronic theory of Bussmann-Holder and Keller [8, 13] reproduces $\alpha_{//}(n)$ quite well near optimum doping, see Fig. 1. In their work, the authors deduced that the *local* polaronic lattice deformation is of $t_2/t_4$, i.e., Jahn–Teller type. From this fact, we may assume that this is also the case in the entire range shown in Fig. 2. In other words, the effect of doping-dependent oxygen isotope and the theories shown in Figs. 1 and 2 are evidence of the planar polaronic origin of the high-temperature superconductivity.

The above conclusion agrees with the ones reached in the viewpoint published by Müller [18] some years ago. In this viewpoint, the possibility for superconductivity to be due to magnetic interactions was also addressed. At that time, no electronic theories were known to yield finite $T_c$'s. Furthermore, experiments were discussed which indicated that magnetic interactions could not be responsible for the high-temperature superconductivity. However, the group of Keimer at the MPI in Stuttgart and others investigated the antiferromagnetic (AFM) peaks near the M-point of the Brillouin zone with inelastic neutron scattering [19] for a number of compounds. These peaks are very weak, and the group was able to enhance the signal by first exciting electrons from the 2*p* level of oxygen with x-rays. The signal obtained in this way quantitatively follows the temperature-dependent superconducting gap. From this, a magnetic mechanism was deduced by assuming an intrinsic homogeneous state of the material. However, as is has been pointed out recently [18] and also earlier [20] this seems not to be the case. For example, X-ray [21] and early Tm NMR [22] investigations yielded near-insulating AFM regions and metallic-like regions forming clusters. Close to a superconducting cluster or stripe, weak AFM signals exist at the border, owing to the continuous electronic part of the wavefunction [23]. Indeed, weak spin ordering has been detected by nuclear quadrupole resonance [24]. It is thus possible that this weak AFM signal is coherent by interaction with the coherent superconducting part, and is responsible for the AFM signal.

The largest oxygen isotope effect at $T_c$ shown in Fig. 1, occurring at low doping, is about $\alpha = 1.0$, i.e., much lower than those observed at the pseudogap temperature $T^*$ reported for LSCO [25] and HoBa$_2$Cu$_4$O$_8$ [26]. For the latter, $\alpha^* = -2.2$. It is sign-inverted, as expected from Eq. (1), which also applies to the pseudogap isotope effect [1]. With the nearly linear $T^*(n)$, we arrive at $\gamma^* = 1.7$ from these data. The existence of an even greater $\alpha^*$ for copper in YBCO in terms of local lattice conformations [26] has been commented on in [8, 18]. In the latter paper, the occurrence of $T^*$ has been assigned to the formation of intersite Jahn–Teller bipolarons [18]. Very recently, the oxygen isotope effects in the LSCO system were studied by means of Cu K-edge XANES, giving another evidence for an enormous involvement of the lattice in the formation of the pseudogap, consistent with a polaronic approach to cuprate superconductivity [27].

The isotope effect with $\alpha = 0.5$ in the classical superconductors was substantial for the acceptance of the BCS theory. The carrier-dependent $\alpha_{//}(n)$ at $T_c(n)$ quantitatively follows the Kresin–Wolf formula [Eq. (1)] derived for polarons along the *c*-direction over the entire doping range [see Fig. 2]. We regard this as a strong clue that, for the materials shown, superconductivity is induced by polarons, or rather bipolarons, with spin $S = 0$, but that they occur *in* the CuO$_2$ plane [18]. The latter follows from the agreement of the calculated planar isotope effects at $T_c(n)$ with the vibronic theory near optimum doping, see Fig. 1. Comparing the observed isotope effects with the vibronic theory indicates that the *local polaronic conformations* are of the $t_2/t_4$ Jahn–Teller type. This theory also yields the large inverted oxygen isotope effect of $\alpha^* = -2.2$ observed at the pseudogap temperature $T^*$ and other behaviors [13]. Furthermore, already at very low doping, intersite Jahn–Teller bipolarons are formed, which cluster in probably ramified, conducting entities, such as stripes [28].

With this, I would like to that Alex Müller for various fruitful discussion and for the perpetual support at the Physics Institute of the University of Zurich. I experienced a pleasant and highly salutary time in Zurich and enjoyed our innumerable colloquies concerning physics and others.

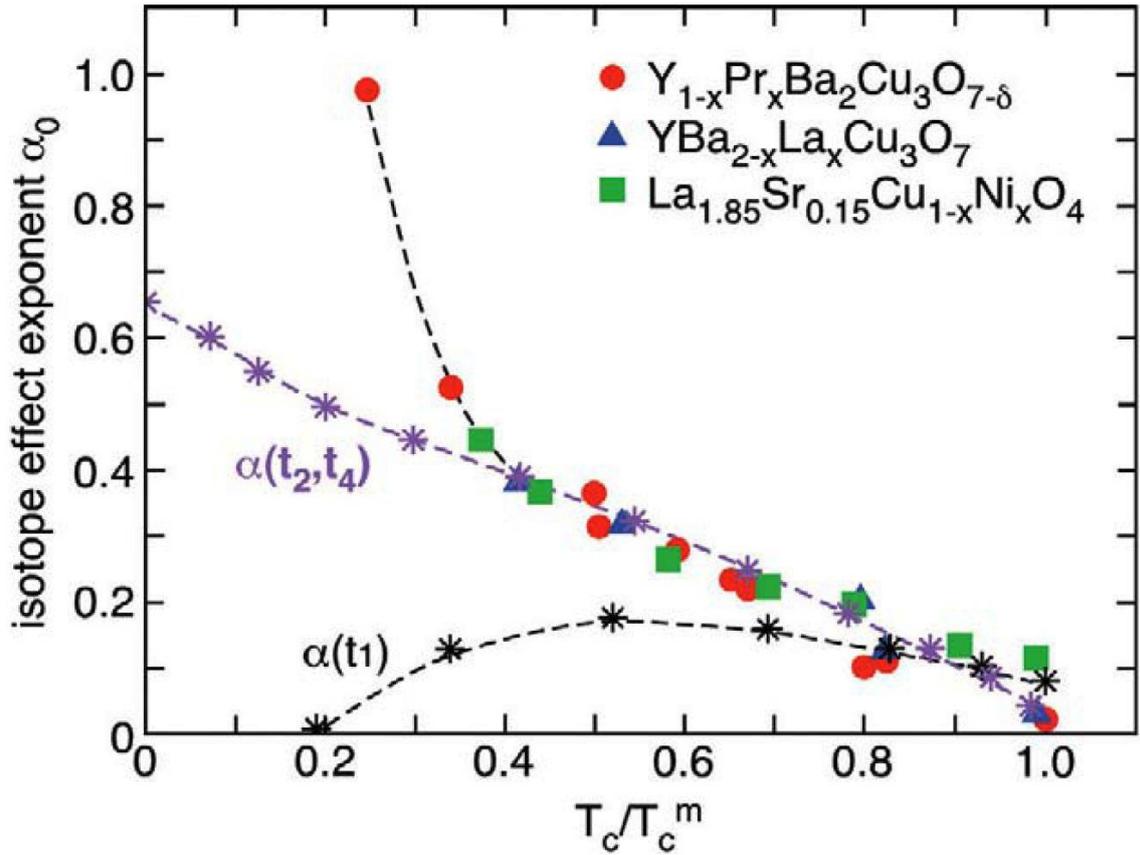

Figure 1: (Color online) The oxygen isotope effect exponent $\alpha$ as a function of $T_c/T_{c,\max}$. The black stars refer to the calculated $\alpha$ when only the nearest-neighbor hopping integral ($t_1$) is renormalized. The purple stars are theoretically derived, with both the second-nearest-neighbor and the interplanar hopping integrals ($t_2, t_4$) renormalized. Red, blue and green data are experimental values for $\alpha$ for $Y_{1-x}Pr_xBa_2Cu_3O_{7-\delta}$ [9, 10], $YBa_{2-x}La_xCu_3O_{7-\delta}$ [11], and $La_{1.85}Sr_{0.15}Cu_{1-x}Ni_xO_4$ [12], respectively. The figure is taken from Ref. [13].

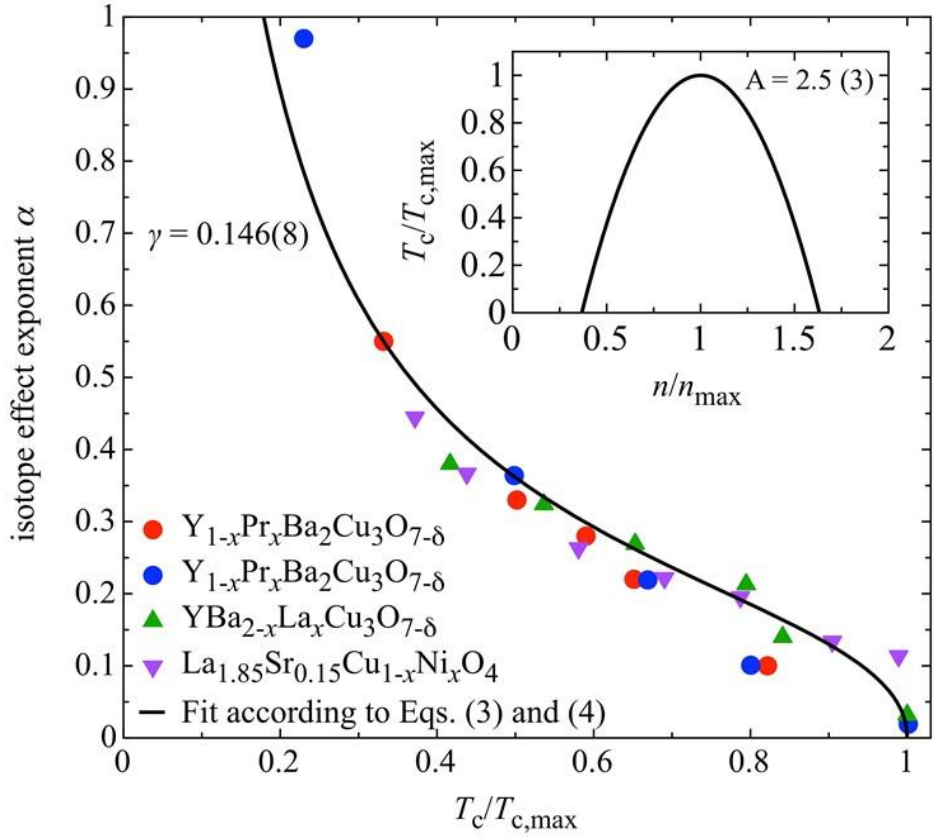

Figure 2: (Color online) The same data for $\alpha$ as in Fig. 1 as a function of $T_c/T_{c,max}$ for $Y_{1-x}Pr_xBa_2Cu_3O_{7-\delta}$ [9, 10], $Y_{1-x}Pr_xBa_2Cu_3O_{7-\delta}$ [11], and $La_{1.85}Sr_{0.15}Cu_{1-x}Ni_xO_4$ [12]. Clearly, Eq. (3) describes the universal dependence of $\alpha$ rather well. We can derive a value of $\gamma = 0.146(8)$ from the fit. The inset shows the doping-dependent $T_c$ according to Eqs. (3) and (4) using the fitted $A = 2.5(3)$. The figure is taken from Ref. [2].

9